# Slow cooling dynamics of the Ising $p$-spin interaction spin-glass model

D. M. Kagan, M. V. Feigelman

Landau Institute for Theoretical Physics, 117940 Moscow, Russia

June 13, 1995*

## Abstract

We have studied dynamical behaviour of the infinite-range Ising spin glass model with $p$-spin interaction above and below the transition into the non-ergodic phase. The transition is continuous at sufficiently high external magnetic field. The dynamic critical exponent of the power-law decay of the autocorrelation function at the transition point is shown to decrease smoothly to zero as the field approaches the "tricritical" point from above; at lower fields the transition is discontinuous. The *slow cooling* approach is used to study the nonergodic behavior below the transition at zero external field. It is shown that the anomalous response function $\Delta(t,t')$ contains $\delta$-function as well as regular contributions at *any* temperature below the phase transition. No evidence of the second phase transition (known to exist within the static replica solution of the same model) is found. At lower enough temperatures the *slow cooling* solution approaches the one known for the standard SK model.

# 1 Introduction

Free energy surface of the spin glasses in the low temperature phase has a very complicated structure and consists of the exponentially large number of valleys with infinite barriers between them (at least in the mean field approximation). This leads to the unique dynamical properties. On the arbitrary long (but finite) time scales such systems occupy only one valley and physical quantities depending only on spin variables at a single time differ from their static values derived from the Gibbs distribution. Usually such a behaviour is referred to as a "nonergodic" one. The term "nonergodicity" means here that the values of measurable physical quantities (magnetization, susceptibility, etc.) cannot be considered just as some functions of the temperature and magnetic field at the measurement point $(T, H)$; rather, they are *functionals of the trajectory* on the $(T, H)$ plane which lead to the final state one is measuring. The quantitative theoretical approach suitable for the study of such an nonergodic behaviour in a classical Sherrington-Kirkpatrick (SK) spin glass [1] and called *slow cooling theory* was invented in the end-80's by Ioffe *et al* [2, 3] (see also [4]) and then developed (in a slightly different version) in [5].

The simplest and most known example of the nonergodic behavior is the difference between so-called zero-field-cooled (ZFC) and field-cooled (FC) susceptibilities ($\chi_{ZFC}$ and $\chi_{FC}$ below), which is also well-described by the static replica-symmetry-breaking approach by Parisi [6, 7]. However, it was shown in [2, 4] that the values of the FC susceptibilities obtained within *slow cooling* approach differ from the Parisi theory results; moreover, the same applies even to the values of the internal energy (which might be considered as a quite robust quantity). Formally one can understand the origin of difference between the results of *slow cooling* and equilibrium theories as being due to non-commutativity of two limiting procedures: the thermodynamic limit $N \to \infty$ and stationary-state limit $t_a \to \infty$ (here $t_a$ is the time system spend in the glassy phase, i.e. *ageing time*). WIthin the *slow cooling* theory it is presumed that the limit $N \to \infty$ is taken first, whereas in the equilibrium theory the limit $t_a \to \infty$ is presumed to be done before it. As a result, the transitions between different valleys of the spin glass phase space which are separated by "infinite" (in the limit $N \to \infty$) barriers are strictly prohibited within the *slow cooling* approach, so the contributions of different valleys do *not* follow Gibbs distribution - contrary to the case of the equilibrium theory.

As a result, there are two sources of the differences between the values of the same physical quantity (e.g. energy) calculated within these two approaches: i) the metastable state (valley) within which the system is typically stuck after *slow cooling* is non-optimal, i.e. has higher free energy than the ground state; ii) in the equilibrium there is an additional contribution from the sum over different valley ( a similar quantity is sometimes called configurational entropy or complexity [8]). It is not quite clear to us which of the above two sources contributes more; however, there is a clear parallel between the temperature dependence of the complexity of the SK model [9]

$$S_{con}(T)/N \propto (T_c - T)^6 . \qquad (1)$$

and the relative difference between *slow cooling* and equilibrium free energies which behaves in the same way near $T_c$ (cf. [2, 3]). Therefore it looks natural to expect that either the second above-mentioned reason (item ii)) is the decisive one, or both of them are of the same importance.

It is seen from 1 that in the vicinity of the spin glass transition (which is the only analytically tractable region) the disagreement between the results of two theories for physical quantities of the SK model is very weak; indeed, relative difference is found to be of the order $(T_c - T)^3$ in the $\chi_{FC}$ and of the order $(T_c - T)^5$ in the internal energy, and did not receive much attention. The same statement applies to other spin-glass models that are characterized by continuous Parisi function $q(x)$ which differs only weakly (at $T$ close to $T_c$) from its maximum value $q(1)$.

There is another family of spin glasses that are characterized by one-step replica symmetry breaking (Parisi function in this models is a step function [17, 10, 11]). Moreover, these models are known to possess dynamic (at $T = T_D$) and static $T = T_c$ phase transitions at different temperatures ($T_D > T_c$). Therefore, one expects all effects of history-dependence and non-ergodicity to be more dramatic in the models of this second family. Note that the configurational entropy is known to be finite in these models right at the transition point:

$$S_{con}(T)/N \propto \Theta(T_D - T), \qquad (2)$$

see [8] and references therein. Indeed, as it will be shown below in the present paper, the *slow cooling* solution for the model of that kind differ *qualitatively* from the results of static replica theory. Another reason to be interested by this second family is the development of glass models *without preexistent disorder* [12, 13] whose behaviour seems to be similar the random spin glasses of the second family. Also similar dynamical equations were derived by Leutheusser [14] in the connection with structural glass problem.

Recently an important progress was achieved in the investigation of the dynamics of spherical $p$-spin interaction model [15], which is exactly solvable in some sense: dynamical equations for this model can be written exactly. In this paper we discuss more complicated (and more realistic) case of the Ising $p$-spin interaction spin-glass model which belongs to the family mentioned above. In the case of infinite $p$ the model is equivalent to random energy model and exhibits one-step replica symmetry breaking [17]. For a finite values of the parameter $p$, this phase is stable in some vicinity of the transition point only, whereas at lower temperatures the second phase transition with a full continuous replica symmetry breaking takes place [10]. This property distinguishes the Ising version of $p$-spin model from the spherical one and it is interesting to inspect how it influences the dynamics.

The paper is organized as follows. In the sections 2 and 3 we introduce the model and derive averaged over disorder mean field generating functional for the correlation and response functions. In the section 4 ergodic dynamics is investigated and transition line $T_D(b)$ (where $b$ is the external magnetic field) is found. It is shown that at high enough magnetic field $b > b_{cr}$ phase transition is of continuous nature and reminds the one known to exist in the SK model at finite field. The dynamic critical exponent $\nu_1$ characterizing the decay of spin-spin autocorrelation function at the dynamic transition line, $C(t) \propto t^{-\nu_1}$, is found as function of $b$, and shown to vanish as $b \to b_{cr} + 0$. At lower external fields the transition is discontinuous: nonzero long-time limit of the autocorrelation function $C(t \to \infty)$ appears just at $T = T_D(b)$, at zero field $C(t \to \infty)|_{T_D} = p - 2$ at small $p - 2$. The "tricritical" field value separating the regions of first- and second-order spin glass transitions is identified as $b_{cr} \propto p - 2$. The reason for an existence of tricritical field $b_{cr}$ is rather simple: in a non-zero external field spins are polarized already at $T > T_D(b)$, i.e. local $< \sigma_i > \neq 0$; the interaction between $\tilde{\sigma} = \sigma_i - < \sigma_i >$ on different sites $i, j$ will now contain usual pair-wise random term $\tilde{\sigma}_i \tilde{\sigma}_j \tilde{J}_{ij}$ with the relative strength $\propto b$, which tends to produce usual SK-type transition and competes with an original $p$-spin interaction. Thus, at high enough fields the "induced" interaction wins and transition is of continuous nature. Phase diagram in the $T - b - p$ coordinates is shown schematically in the figure 1.

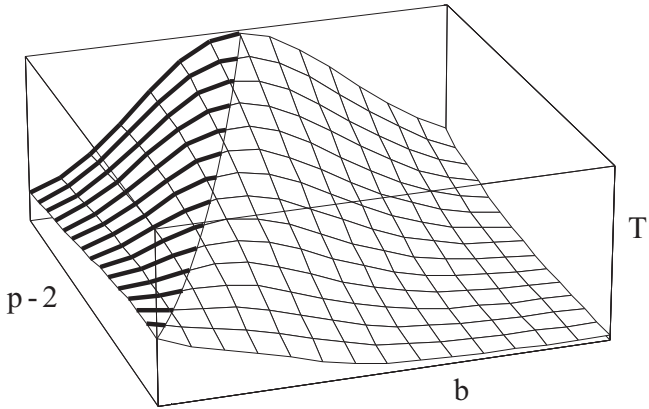

Figure 1: Phase diagram. Thin lines: continuous transition, thick lines: discontinuous transition.

In the section 5 slow cooling of the system in the spin-glass phase is considered. It is shown that the behaviour of the nonergodic dynamic response is qualitatively the same in the whole low-temperature phase, i.e. there is no additional low-$T$ phase transition known from the replica solution [10]. As the temperature goes down, the solution for the anomalous correlation and response functions interpolate smoothly between the one characteristic for spherical $p > 2$-spin model [15] and the one for the usual $p = 2$ Ising model [2]. We also found the *downward jump* in the specific heat as temperature decreases through $T = T_D$. Such a behaviour is known to exist in real glasses; in our model its origin may be associated with an abrupt drop in the configurational entropy (cf. 2) due to the freezing of the system in the one of all possible ($\propto e^{S_{com}}$) metastable states.

Section 6 is devoted to the discussion of the results.

## 2 The model

Ising $p$-spin interaction spin-glass model is described by Hamiltonian [17, 10]:

$$H = - \sum_{1 \leq i_1 < \cdots < i_p \leq N} J_{i_1 \ldots i_p} \sigma_{i_1} \cdots \sigma_{i_p} - b \sum_{i=1}^{N} \sigma_i \qquad (3)$$

where $b$ is an external field. The spin glass, described by this Hamiltonian, is a system of $N$ Ising spin interacting via randomly quenched infinite range interactions $J_{i_1...i_p}$. We will take, for simplicity, distribution of constants $J_{i_1...i_p}$ to be Gaussian :

$$P(J_{i_1...i_p}) = \left(\frac{N_{p-1}}{\pi J^2 p!}\right)^{1/2} exp\left[-\frac{J_{i_1...i_p}^2 N_{p-1}}{J^2 p!}\right], \ 1 \leq i_1 < \cdots < i_p \leq N \ . \quad (4)$$

We will assume Glauber dynamics for $\sigma_i$: the probability for $\sigma_i$ to change it's sign during unit time $\Gamma^{-1}$ is

$$w\{(\sigma_1,\ldots,\sigma_i,\ldots,\sigma_N) \to (\sigma_1,\ldots,-\sigma_i,\ldots,\sigma_N)\} = \frac{1}{2}(1 - \sigma_i \tanh \beta h_i(t)) \quad (5)$$

where local field $h_i = \dfrac{\partial H}{\partial \sigma_i}$, $\beta = T^{-1}$ is inverse temperature. In the following, we will put $\Gamma = 1$ without loss of generality.

The quantities of our interest are the average response function

$$G(t,t') = \overline{\frac{\delta \langle \sigma_i(t) \rangle}{\delta \beta b_i(t')}}$$

that vanishes for $t < t'$, and the average spin correlation function

$$C(t,t') = \overline{\langle \sigma_i(t)\sigma_i(t') \rangle}$$

Here $\langle \ldots \rangle$ means the dynamic average and $\overline{\ldots}$ — the average over disorder.

## 3   Averaged self-consistent generating functional

In the thermodynamic limit $N \to \infty$, the analysis simplifies and dynamics of the system can be described by a set of self-consistent equations for a single spin. These equations can be derived by introducing a generating functional for Glauber dynamics [19, 20] and averaging it over disorder. This functional is defined as $Z[\hat{\lambda}] = \left\langle exp\left(\sum_{i=1}^{N} \int \hat{\lambda}_i(t)\sigma_i(t)dt\right)\right\rangle$ and, as shown in Appendix, can be written as:

$$Z = \hat{J}Z_0\Big|_{h=\beta b} \ .$$

Here
$$\hat{J} = exp\left(\sum_{i_1<i_2<\cdots<i_p} \int dt J_{i_1\ldots i_p}\hat{\delta}_{i_1}(t)\hat{\delta}_{i_2}(t)\cdots\hat{\delta}_{i_p}(t)\right),$$

$$\hat{\delta}_t = \frac{\delta}{\delta\hat{\lambda}(t)}, \quad \delta_t = \frac{\delta}{\delta h(t)},$$

$Z_0[\hat{\lambda}]$ is the generating functional for non-interacting spins:

$$Z_0[\hat{\lambda}] = exp\left(\sum_{i=1}^{N}\int_{t_0}^{\infty}\hat{\lambda}_i(t)m_i(t)dt\right),$$

and $m_i$ obey the equation:

$$\partial_t m_i(t) = i\hat{\lambda}_i(t)(1 - m_i^2(t)) - (m_i(t) - \tanh h), \quad m_i(t_0) = m_0. \qquad (6)$$

It is convenient to use this functional in another form. Note that

$$Z_0[\hat{\lambda}] = \int \prod_{i=1}^{N} \mathcal{D}\hat{\sigma}_i \mathcal{D}\sigma_i \mathcal{D}h_i \mathcal{D}\hat{h}_i \, exp\left(\sum_{i=1}^{N}\int \hat{\lambda}_i \sigma_i dt\right) exp(S),$$

where

$$S = \sum_i \left\{-i\int \hat{\sigma}_i(\sigma_i - m_i)dt - i\int \hat{h}_i(h_i - \beta b)\, dt\right\},$$

$$\partial_t m_i(t) = i\hat{\sigma}_i(t)(1 - m_i^2(t)) - (m_i(t) - \tanh\beta b), \quad m_i(t_0) = m_0.$$

Action of operator $\hat{J}$ gives

$$Z[\hat{\lambda}] = \int \prod_{i=1}^{N} \mathcal{D}\hat{\sigma}_i \mathcal{D}\sigma_i \mathcal{D}h_i \mathcal{D}\hat{h}_i \, exp\left(\sum_{i=1}^{N}\int \hat{\lambda}_i \sigma_i dt\right) exp(S) \qquad (7)$$

where

$$S = \sum_i \left\{-i\int \hat{\sigma}_i(\sigma_i - m_i)dt - i\int \hat{h}_i\left(h_i - \beta\frac{\partial H}{\partial \sigma_i}\right)dt\right\}.$$

This integral is normalized to unity (i.e. $Z[\hat{\lambda} \equiv 0] = 1$), so that average over $J_{i_1\ldots i_p}$ can be performed. The following derivation of the mean field generating functional is standard and almost coincides with the analogous one derived by Kirkpatrick and Thirumalai [11] for the case of Langevin dynamics. There is no need to report this derivation and we write only ultimate result:

$$Z[\hat{\lambda}] = \int \mathcal{D}h\mathcal{D}\hat{h}\mathcal{D}\sigma\mathcal{D}\hat{\sigma}\, exp(S) \tag{8}$$

$$S = -i\int \hat{h}(h - \beta b) - \int [i\hat{\sigma}(\sigma - m) - \hat{\lambda}\sigma] + S_{gl}, \tag{9}$$

$$S_{gl} = -J^2 \int \frac{1}{2}\sqrt{\mu(1)\mu(2)}C^{p-1}(1,2)\hat{h}(1)\hat{h}(2) +$$

$$+ J^2 \int (p-1)\sqrt{\mu(1)\mu(2)}G(1,2)C^{p-2}(1,2)i\hat{h}(1)\sigma(2), \tag{10}$$

$$\partial_t m(t) = i\hat{\sigma}(t)(1 - m^2(t)) - (m(t) - \tanh h),\; m(t_0) = m_0. \tag{11}$$

with the obvious abbreviation of time arguments. Here we use the notation $\mu = p\beta^2 J^2/2$. The correlation and response functions $C(t,t')$ and $G(t,t')$ have to be determined self-consistently from (8) with

$$C(t,t') = \frac{\delta}{\delta\hat{\lambda}_t}\frac{\delta}{\delta\hat{\lambda}_{t'}}Z[\hat{\lambda}]\Big|_{\hat{\lambda}\equiv 0}$$

and

$$G(t,t') = \frac{\delta}{\delta\hat{\lambda}_t}\frac{\delta}{\delta\beta b_{t'}}Z[\hat{\lambda}]\Big|_{\hat{\lambda}\equiv 0}.$$

In the case of $p = 2$ the action in (8) coincides with the action of Sommers [19] for the Sherrington-Kirkpatrick model. Further we will be mainly interesting in the case of $\varepsilon = p - 2 \ll 1$, since we already can consider $p$ as a continuous variable. We will put also $J = 1$ to simplify expressions.

## 4 Dynamics above the transition line

In this section we study the dynamical mean field equation for constant temperature and magnetic field. Under these conditions, the system is expected to be in the paramagnetic phase (see the criterion below), so that the correlation and response functions depend only on time difference and are related by the fluctuation-dissipation theorem (FDT)

$$G(t) = -\theta(t)\partial_t C(t).$$

Here it is supposed that $t_0$ from (6) is equal to $-\infty$, so that $m$ does not depend on $m_0$. In the non-zero magnetic field, Edwards-Anderson order parameter $q = \lim_{t \to \infty} C(t)$ is not equal to zero, too, at any temperature, so it is convenient to represent the correlator $C(t)$ as $C(t) = \tilde{C}(t) + q$. The part of $S_{gl}$ containing $q$ is

$$exp\left(-\int \frac{1}{2}\mu q^{p-1} \hat{h}(1)\hat{h}(2) d1 d2\right) = \left\langle exp\left(iz \int \hat{h}(t) dt\right)\right\rangle_z$$

where $\langle \ldots \rangle_z$ means $\int \frac{dz}{\sqrt{2\pi \mu q^{p-1}}}(\ldots) \exp\left(-\frac{z^2}{2\mu q^{p-1}}\right)$.

Analogous to previous section, one can write $Z[\hat{\lambda}]$ in the following form:

$$Z = \left\langle \hat{J} Z_0 \Big|_{h=\beta b + z}\right\rangle_z$$

where $\hat{J}$ is

$$\hat{J} = exp\left[\frac{1}{2}\int \hat{C}(1,2)\frac{\delta}{\delta h(1)}\frac{\delta}{\delta h(2)} + \int \hat{G}(1,2)\frac{\delta}{\delta h(1)}\frac{\delta}{\delta \hat{\lambda}(2)}\right] \quad (12)$$

and

$$\begin{aligned}
\hat{C}(1,2) &= \sqrt{\mu(1)\mu(2)}\ C^{p-1}(1,2) - \sqrt{\mu(1)\mu(2)}\ q^{p-1} \\
\hat{G}(1,2) &= (p-1)\sqrt{\mu(1)\mu(2)}\ C^{p-2}(1,2) G(1,2).
\end{aligned}$$

The function $m$ in $Z_0$ obeys equation (6).

One can show [19, 20] by means of FDT that the static limits of all correlation functions are independent of the short-time parts of $C$ and $G$ in (12). This leads to the equation

$$q = \left\langle m^2 \right\rangle \quad (13)$$

where $m = \tanh(z + \beta b)$ and $\langle \ldots \rangle$ means average over $z$. This equation coincides with the replica-symmetric equation, found in [17, 10].

Moreover, we will see that for $t \gg 1$ the expansion of $C$ and $G$ in $(\beta J)$ from (12) is at the same time an expansion in $\hat{G}(\omega) - G(\omega)|_{\omega=0}$. Therefore, one is able to use perturbation theory with respect to $\hat{C}$ and $\hat{G}$.

Following Sommers [19], we write several terms of this expansion applying FDT to each term:

$$G(\omega) = \frac{\langle 1 - m^2 \rangle}{1 - i\omega} + \frac{i\omega \langle (1-m^2)^2 \rangle}{(1-i\omega)^2} \left[ \int_0^\infty e^{i\omega t} \hat{C}(t) dt + \hat{C}(t)|_{t=0} \right] -$$

$$\frac{\omega^2}{(1-i\omega)^3} \langle (1-m^2)^3 \rangle \left[ \int_0^\infty e^{i\omega t} \hat{C}(t) dt + \hat{C}(t)|_{t=0} \right]^2 +$$

$$\frac{i\omega}{(1-i\omega)^2} \langle 2m^2(1-m^2)^2 \rangle \left[ \int_0^\infty e^{i\omega t} \left(\hat{C}(t)\right)^2 dt + \left(\hat{C}(t)|_{t=0}\right)^2 \right] \quad (14)$$

Let us consider, at first, sufficiently large magnetic fields where the transition is expected to be of the second order (the exact criterion will be derived below). The correlator $C(\omega)$ at $\omega \to 0$ diverges at the transition line where

$$\mu_c(p-1) \langle (1-m^2)^2 \rangle q_c^{p-2} = 1 \ . \quad (15)$$

At $T = T_c$ we have $C(t) = At^{-\nu_1}$ where $\nu_1$ obeys the equation:

$$\pi \cot \pi \nu_1 \langle (1-m^2)^3 \rangle \frac{\Gamma(2\nu_1)}{\Gamma^2(\nu_1)} = \langle m^2(1-m^2)^2 \rangle +$$
$$\langle (1-m^2)^2 \rangle \frac{p-2}{4\mu q^{p-1}(p-1)} \ . \quad (16)$$

For the case of $p = 2$, this equation was derived in [18, 19]. On the contrary to Sherrington-Kirkpatrick model with $p = 2$, $\nu_1$ becomes zero at tricritical point where

$$\langle (1-m^2)^3 \rangle = \langle 2m^2(1-m^2)^2 \rangle + \langle (1-m^2)^2 \rangle \frac{p-2}{2\mu q^{p-1}(p-1)} \ . \quad (17)$$

This equation along with equations (13) and (15) determines $T_{tr}$, $q_{tr}$ and $b_{tr}$. Near tricritical point we can expand left hand side of (16) over $\nu_1$. This expansion does not contain first order term and begins with $\nu_1^2$, so that $\nu_1$ at $b - b_{tr} \ll 1$ is:

$$\nu_1 \propto \sqrt{b - b_{tr}} \ . \quad (18)$$

We can also find how $b_{tr}$ tends to zero at the limit $p \to 2$. The expansion of the equations (13), (15) and (17) over $m^2$ yields

$$b_{tr} \propto \varepsilon = p - 2 \ . \quad (19)$$

At the field $b < b_{tr}$ and $T = T_c$ the correlation function $C(t)$ does not vanish in the limit $t \to \infty$ what makes phase transition to be discontinuous. Let us consider the case of $b = 0$. At $\omega \ll 1$

$$G(\omega) = 1 + i\omega \int_0^\infty e^{i\omega t} \hat{C}^{p-1}(t) dt + \left[ i\omega \int_0^\infty e^{i\omega t} \hat{C}^{p-1}(t) dt \right]^2 . \qquad (20)$$

All results at $b = 0$ will be correct only for $\varepsilon = p - 2 \ll 1$, since we restrict ourselves to several terms of expansion of (12). Let's assume that $C(t)$ have time-persistent part:

$$\lim_{t \to \infty} C(t) = q$$

Equation (20) yields (in the main order of $q$)

$$q - \mu q^{1+\varepsilon} + q^2 = 0$$

This equation has trivial solution $q = 0$ and nontrivial one starting from some $\mu = \mu_c$. This condition determine transition point:

$$\begin{aligned} q_c &= \varepsilon, \\ \mu_c &= 1 - \varepsilon \ln \varepsilon + \varepsilon. \end{aligned} \qquad (21)$$

These results will be found below by adiabatic cooling method. It should be emphasized that the transition temperature is higher than the one derived in the [10] with the help of replica method.

To determine critical behaviour of function $C(t)$ at long times let us consider it's Laplace transform $C(\omega) = \int_0^\infty e^{i\omega t} C(t) dt$. We suppose that at $T$ slightly above $T_c$ function $C(\omega)$ have a pole contribution and a remaining part:

$$C(\omega) = \frac{q}{1/\tau - i\omega} + \widetilde{C}(\omega) , \qquad (22)$$

where relaxation time $\tau$ diverges when $T \to T_c$. At the transition point $\widetilde{C}(\omega) \propto \omega^{\nu_2 - 1}$ (and $\widetilde{C}(t) \propto t^{-\nu_2}$). To determine $\nu_2$ let us consider equation (20) at $T = T_c$. The terms of the order of $\omega^{\nu_2}$ compensate each other in agreement with (21). The comparing of the coefficients in terms of the order of $\omega^{2\nu_2}$ gives the equation for $\nu_2$. In the main order in $\varepsilon$

$$\Gamma(1 - 2\nu_2) = 2\Gamma^2(1 - \nu_2), \quad \nu_2 \approx 0.395 . \qquad (23)$$

We can also find how relaxation time $\tau$ depends on the temperature $T$ near $T_c$. Equation (20) at $\omega = 0$ yields

$$1 = (1+\varepsilon)\mu q^{\varepsilon} - 2q$$

As will be shown below characteristic time $\tilde{\tau}$ of the function $\tilde{C}(t)$ is much smaller than $\tau$. At the interval $1/\tau \ll \omega \ll 1/\tilde{\tau}$ one can obtain in the main order of $\varepsilon$:

$$0 = \mu_c - \mu + \frac{2\varepsilon}{\tau}\tilde{C}(\omega) + \frac{i\omega\varepsilon}{\tau}\frac{d}{d\omega}\tilde{C}(\omega) + \left(i\omega\tilde{C}(\omega)\right)^2 + \frac{i\omega}{2}\int_0^{\infty} e^{i\omega t}\tilde{C}^2(t)dt \ .$$

Such an equation (at $\varepsilon = 1$) was studied by Leutheusser [14] in his description of the dynamics near the liquid-glass transition. Asuming that $\tilde{\tau} \propto (\mu_c - \mu)^{-\alpha}$ and $\tau \propto (\mu_c - \mu)^{-\beta}$ one obtains

$$\begin{aligned} \alpha &= \frac{1}{2\nu_2} \\ \beta &= \frac{1+\nu_2}{2\nu_2} \end{aligned} \qquad (24)$$

It should be mentioned that for the spherical $p$-spin interaction model [15] the response function is

$$G(\omega) = \left(1 - i\omega \int_0^{\infty} e^{i\omega t}\hat{C}^{p-1}(t)dt\right)^{-1} \ . \qquad (25)$$

For small $\varepsilon$, this expression is identical to (20). Therefore, the asymptotic behaviour of the functions $C(t)$ and $G(t)$ is the same as for the spherical model (at least in the main order of $\varepsilon$), as one can see at equations (21), (23) and (24).

## 5 Slow cooling

### 5.1 Adiabatic equations and transition line

Probably the most direct way to investigate the behaviour of the spin glass on a finite time scale is based on slow cooling starting in the ergodic high temperature phase. We will assume that temperature and, maybe, magnetic field vary on a time scale of the order of $t_0 \gg \Gamma^{-1}$. The situation is sufficiently complicated because the correlation and response functions depend now on both time arguments and no longer on the time difference. It is convenient to divide this functions into "fast" and "slow" parts:

$$C(t,t') = \widetilde{C}_t(t-t') + q(t,t') \;,$$

$$G(t,t') = \widetilde{G}_t(t-t') + \Delta(t,t') \;.$$

The functions $\widetilde{C}_t(t-t')$ and $\widetilde{G}_t(t-t')$ decay on the time scale $\bar{t} \ll t_0$ and represent the dynamics within one of the "pure" states in the system phase space. The relevant time scale of the functions $q(t,t')$ and $\Delta(t,t')$ is $t_0$. It turns out that it is possible to obtain the closed system of equations for the "slow" parts of the correlation and response functions. For the first time it was proposed by Ioffe at all in [3, 2] where the Langevin dynamics was used. Alternative method which starts from Glauber dynamics was developed by Horner and Freixa-Pascual in [5]. It can be proved (see Appendix 2) that both methods give identical equations. Here we prefer to use the second one.

Now let us explain briefly the main idea. As it was mentioned in the previous section, the functions $q$ and $\Delta$ at $t - t' \gg \bar{t}$ are independent of the short time functions $\widetilde{C}_t(t-t')$ and $\widetilde{G}_t(t-t')$. Thus we can replace $C$ and $G$ by $q$ and $\Delta$ respectively in the functional (12). Moreover, we also can consider $m(t)$ to be equal to $\tanh(\beta b)$. The existence of the terms $\partial_t m(t)$ and $i\hat{\sigma}(t)(1 - m^2(t))$ in the equation (11) leads to the fact that correlation functions are not equal to the asymptotic values and have also relaxation parts. For example, the correlator $C(t) = \langle \sigma(t)\sigma(0) \rangle$ of *free* spins in the constant field is $C(t) = \tanh^2(\beta b) + (1 - \tanh^2(\beta b))e^{-|t|}$. Without these terms, we would obtain $C(t) = \tanh^2(\beta b)$. Actual form of the time-dependent correlation function for the interacting spins is, of course, different from simple exponential relaxation; however, its particular form is irrelevant for the derivation of the *slow cooling* equations for the slow functions $\Delta(t,t')$ and $q(q,q')$.

If we take into account this remarks, the generating functional (12) in the adiabatic limit can be written as

$$Z[\hat{\lambda}] = \hat{J} exp \left( \int \hat{\lambda}(t) \tanh h(t) dt \right) \bigg|_{h=\beta b}, \qquad (26)$$

$$\hat{J} = exp\left[\frac{1}{2}\int \sqrt{\mu(1)\mu(2)}q^{p-1}(1,2)\frac{\delta}{\delta h(1)}\frac{\delta}{\delta h(2)} + \right.$$
$$\left. (p-1)\int \sqrt{\mu(1)\mu(2)}\Delta(1,2)q^{p-2}(1,2)\frac{\delta}{\delta h(1)}\frac{\delta}{\delta\hat{\lambda}(2)}\right] . \tag{27}$$

As in the previous section, we introduce the functions

$$\hat{q}(t,t') = \sqrt{\mu(t)\mu(t')}q^{p-1}(t,t') - \mu_c q^{p-1}|_{T=T_c+0} ,$$

$$\hat{\Delta}(t,t') = (p-1)\sqrt{\mu(t)\mu(t')}q^{p-2}(t,t')\Delta(t,t')$$

and rewrite (26) as

$$Z[\hat{\lambda}] = \left\langle \hat{J}exp\left(\int \hat{\lambda}(t)\tanh h(t)dt\right)\bigg|_{h=\beta b+z}\right\rangle , \tag{28}$$

$$\hat{J} = exp\left[\frac{1}{2}\int \hat{q}(1,2)\frac{\delta}{\delta h(1)}\frac{\delta}{\delta h(2)} + \int \hat{\Delta}(1,2)\frac{\delta}{\delta h(1)}\frac{\delta}{\delta\hat{\lambda}(2)}\right] , \tag{29}$$

$$\left\langle z^2\right\rangle = \mu_c q^{p-1}|_{T=T_c+0} .$$

The self-consistence conditions are

$$q(t,t') = \hat{\delta}_t\hat{\delta}'_t\ Z[\hat{\lambda}]\ , \Delta(t,t') = \hat{\delta}_t\delta'_t\ Z[\hat{\lambda}]\ ,$$

where

$$\hat{\delta}_t = \frac{\delta}{\delta\hat{\lambda}(t)},\ \delta_t = \frac{\delta}{\delta\beta b(t)}.$$

To derive equations mentioned above we have to expand $\hat{J}$ (29) and leave several terms of this expansion. This method is correct near the transition point at $b = 0$ and $\varepsilon \ll 1$ and also at $b$ near or greater than $b_{tr}$ and arbitrary $\varepsilon$. At this conditions, the values of $q_{t,t'}$ and $\int \Delta_{t,t'}\,dt'$ is small. The expansion of $\hat{J}$ to the second order in $q$ and $\Delta$ results after some algebra in equations (30) and (31). The essential remark should be done before we write this equations. After the action of operator $\int \hat{\Delta}(1,2)\delta_1\hat{\delta}_2$ on $exp\left(\int \hat{\lambda}(t)\tanh\beta h(t)dt\right)$ it appears the integral $\int \Delta(t,t)(1 - \tanh^2(\beta b))\,dt$ containing undefined value $\Delta(t,t)$. Let us consider what such terms in the exact (not adiabatic) correlators correspond to. To derive corresponding terms the action of $\frac{\delta}{\delta\hat{\lambda}}$ should be extended only on the function $\hat{\lambda}(t)$ in the exponent. We obtain as a result $\int\limits_{t'>t} \Delta(t,t') exp(t-t')\ dt dt'$ . This integral equals zero for any value $\Delta(t,t')$. Therefore, in the adiabatic limit we should set $\Delta(t,t')$ equal to zero.

Thus, the equations for $\Delta$ mentioned above is

$$\Delta_{t,t'} = \left\langle m'^{\,2} \right\rangle \hat{\Delta}_{t,t'} + \left\langle m''^{\,2} \right\rangle + \frac{1}{2}\hat{\Delta}_{t,t'}(\hat{q}_t + \hat{q}_{t'}) \left\langle m'm''' \right\rangle +$$

$$+ \left\langle m'^{\,3} \right\rangle \int \hat{\Delta}_{t,1}\hat{\Delta}_{1,t'}\ d1 + \left\langle m\,m'm'' \right\rangle \hat{\Delta}_{t,t'} \int (\hat{\Delta}_{t,1} + \hat{\Delta}_{t',1})\ d1\ . \qquad (30)$$

The equation for $q$ is

$$q_{t,t'} = \left\langle m'^{\,2} \right\rangle \hat{q}_{t,t'} + \frac{1}{2} \left\langle m\,m'' \right\rangle (\hat{q}_t + \hat{q}_{t'}) + \left\langle m^2 m' \right\rangle \int (\hat{\Delta}_{t',1} + \hat{\Delta}_{t'1})\ d1 +$$

$$+ \left\langle m''^{\,2} \right\rangle \left[ \frac{1}{2}\hat{q}^2_{t,t'} + \frac{1}{4}\hat{q}_t\,\hat{q}_{t'} \right] + \frac{1}{2} \left\langle m'm''' \right\rangle \hat{q}_{t,t'}(\hat{q}_t + \hat{q}_{t'}) + \frac{1}{8} \left\langle m\,m^{(4)} \right\rangle (\hat{q}^2_t + \hat{q}^2_{t'}) +$$

$$+ \left\langle m'^{\,3} \right\rangle \int (\hat{\Delta}_{t,1}\hat{q}_{t',1} + \hat{\Delta}_{t',1}\hat{q}_{t,1})\ d1 + \frac{1}{2} \left\langle m'''m^2 \right\rangle \int (\hat{\Delta}_{t,1}\hat{q}_t + \hat{\Delta}_{t',1}\hat{q}_{t'})\ d1 +$$

$$+ \left\langle m\,m'm'' \right\rangle \left\{ \frac{1}{2}\int [\hat{\Delta}_{t,1}(\hat{q}_{t'} + \hat{q}_1) + \hat{\Delta}_{t',1}(\hat{q}_t + \hat{q}_1)]\ d1 + \right.$$

$$\left. + \int (\hat{\Delta}_{t,1}\hat{q}_{t,1} + \hat{\Delta}_{t',1}\hat{q}_{t',1})\ d1 + \int \hat{q}_{t,t'}(\hat{\Delta}_{t,1} + \hat{\Delta}_{t',1})\ d1 \right\} +$$

$$+ \frac{1}{2} \left\langle m''m^3 \right\rangle \int (\hat{\Delta}_{t,1}\hat{\Delta}_{t,2} + \hat{\Delta}_{t',1}\hat{\Delta}_{t',2})\ d1 d2 +$$

$$+ \left\langle m'^{\,2}m^2 \right\rangle \left[ \hat{\Delta}_{t,1}\hat{\Delta}_{t',2}\ d1 d2 + \int (\hat{\Delta}_{t,1}\hat{\Delta}_{1,2} + \hat{\Delta}_{t',1}\hat{\Delta}_{1,2}) \right]\ . \qquad (31)$$

Here

$$m^n = \frac{d^n}{dz^n} \tanh(z + \beta b)\ .$$

Several remarks should be done about the behaviour of the functions $q(t,t')$ and $\Delta(t,t')$. If the parameter $q$ varies smoothly on the cooling trajectory, these functions vary also smoothly in the time scales of the order of $t_0$. If $q$ jumps discontinuously at the critical temperature, we should expect also discontinuous jump in the function $q(t,t')$ at $t' \sim t_c$. This means that the correlation function $C(t,t')$ varies at $t' \sim t_c$ in the time scales less than the characteristic cooling time $t_0$. Thus the functions $q(t,t')$ and $\Delta(t,t')$ at $t'$ near $t_c$ should be related by generalized FDT:

$$\partial_{t'} q(t,t') = \Delta(t,t') \qquad (32)$$

and consequently

$$\partial_{t'} \hat{q}(t,t') = \hat{\Delta}(t,t')$$

what is consistent with the equations (30) and (31). This condition should be also satisfied if $t - t' \ll t_0$. The same assumptions were used in [15] where the dynamics of the spherical model considered. In another words, the functions $G(t,t')$ and $C(t,t')$ at $t - t' \ll t_0$ and $t - t' \ll t_c$ can be considered like $\delta$-function and $\theta$-function respectively.

Now we show how one can obtain with the help of this generalized FDT and equations (30) and (31) some of the results of the previous section. Let us find, for example, tricritical point. Suppose that the cooling trajectory cross transition line near this point at $b < b_{tr}$. If we put $t = t' = t_c + 0$

$$q = \langle m^2 \rangle + \langle m'^2 \rangle \hat{q} + \hat{q}^2 \langle (1-m^2)^2 (3m^2 - 1) \rangle . \qquad (33)$$

Above $T_c$ $q$ satisfied equation $q = \langle m^2 \rangle$ and at $b = b_{tr}$ jump of the order parameter $\delta q$ becomes zero. Expansion of (33) in $\delta q$ gives in the first order marginal stability condition (15) and in the second order — equation (17).

Consider now the case of $b = 0$ and $t = t' = t_c + 0$. The equations will be

$$1 = (p-1)\mu q^{p-2}(1 - 2\hat{q})$$
$$q = \hat{q}(1 - 2\hat{q}) + \hat{q}^2$$

This yields $q_c = \varepsilon$, $\mu_c = 1 + \varepsilon - \varepsilon \log \varepsilon$ as in the above (21).

## 5.2 Anomalous response function at $b = 0$

Let us now consider the solution for $q$ and $\Delta$ in the spin-glass state below the transition point but at zero external field $b$. The equations (30) and (31) reduce to

$$\begin{aligned}
\Delta_{t,t'} &= \hat{\Delta}_{t,t'} \left(1 - \hat{q}_t - \hat{q}_{t'}\right) + \int \hat{\Delta}_{t,1} \hat{\Delta}_{1,t'} \, d1 , \\
q_{t,t'} &= \hat{q}_{t,t'} \left(1 - \hat{q}_t - \hat{q}_{t'}\right) + \hat{q}_{t,t_c} \hat{q}_{t',t_c} + \int (\hat{\Delta}_{t,1} \hat{q}_{t',1} + \hat{\Delta}_{t',1} \hat{q}_{t,1}) \, d1 \quad (34)
\end{aligned}$$

Here and further integration interval does not contain $t_c$. The solution at $t \geq t'$, $\mu = \mu_c + 2\tau$ and $\tau \ll \varepsilon \ll 1$ is

$$q(t,t') = (\varepsilon + \tau + \tau')\theta(t-t'), \quad \Delta(t,t') = (\varepsilon + \tau)\delta(t'-t_c) \qquad (35)$$

At $\tau, \tau' \gg \varepsilon$

$$q(t,t') = \sqrt{\tau\tau'}\,\theta(t-t'), \quad \Delta(t,t') = \sqrt{\tau\varepsilon}\,\delta(t-t_c), \quad q(t,t_c) = \sqrt{\tau\varepsilon}\ . \qquad (36)$$

As one can see, anomalous response function $\Delta$ consists of the $\delta$-function contribution only. Such a solution, as will be discussed below, is in agreement with the one-step replica simmetry breaking solution obtained previously for the same model within the static approach [17, 10]. However, in turns out that these results holds only approximately and more accurate calculations lead to an appearance of the regular (smooth) part in the anomalous response function.

To find smooth part of anomalous response function $\Delta$ we should expand $\hat{J}$ in (29) up to third order in $q$ and $\Delta$. We also can put $\Delta \propto \delta(t-t_c)$ in the third order terms. Thus, equation for $\Delta$:

$$\begin{aligned}\Delta_{t,t'} &= \hat{\Delta}_{t,t'}\left[1 - \hat{q}_t - \hat{q}_{t'} + 2(\hat{q}_t^2 + \hat{q}_{t'}^2 + \hat{q}_{t,t'}^2) + \hat{q}_t\hat{q}_{t'} - \right.\\ &\quad \left. \hat{q}_{t,t_c}^2 - \hat{q}_{t',t_c}^2 - 2\hat{q}_{t,t_c}\hat{q}_{t',t_c}\right] + \int \hat{\Delta}_{t,1}\hat{\Delta}_{1,t'}\,d1\ .\end{aligned} \qquad (37)$$

Equation for $q$:

$$q_{t,t'} = \hat{q}_{t,t'}\left[1 - \hat{q}_t - \hat{q}_{t'} + 2(\hat{q}_t^2 + \hat{q}_{t'}^2 + \frac{1}{3}\hat{q}_{t,t'}^2) + \hat{q}_t\hat{q}_{t'}\right] + \hat{q}_{t,t_c}\hat{q}_{t,t_c} -$$
$$(\hat{q}_t^2 + \hat{q}_{t'}^2)(\hat{q}_{t,t_c}^2 + \hat{q}_{t',t_c}^2) - \hat{q}_{t,t_c}\hat{q}_{t,t_c}\hat{q}_{t,t'} + \int (\hat{q}_{t,1}\hat{\Delta}_{t',1} + \hat{q}_{t',1}\hat{\Delta}_{t,1})\,d1\ . \qquad (38)$$

Consider region $\tau \ll \varepsilon$ and introduce notation $q(t,t') = \varepsilon + \tau + \tau' + w(t,t')$. Then

$$\varepsilon\Delta(t,t')w(t,t') + \int \Delta_{t,1}\Delta_{1,t'}\,d1 = 0\ ,$$
$$\frac{1}{2}w^2(t,t') + \int (w_{t,1}\Delta_{t',1} + w_{t',1}\Delta_{t,1})\,d1 + \frac{10}{3}\tau\tau' = 0\ . \qquad (39)$$

The solution of this equation has a degeneracy: we can change sign of functions $\Delta$ and $w$ simultaneously. However, we should choose positive value of $\Delta$ since it corresponds to higher magnetization:

$$\Delta(t,t') = (\varepsilon+\tau)\delta(t'-t_c) + \frac{10}{3}\sqrt{\varepsilon}\,\frac{d\tau'}{dt'}\theta(t-t'),\quad w(t,t') = -\frac{10}{3}|\tau-\tau'|\sqrt{\varepsilon}\ . \quad (40)$$

The above choice of solution is based on the following physical arguments: the very presence of the anomalous response function is [2] due to the possibility for the system to choose between different configuration-space valleys (which come into existence during the cooling) in order to minimize the free energy of the final state. In particular, when cooling procedure is done at some constant magnetic field, the valleys with higher magnetization along the applied field direction are certainly preferable — which means that irreversible contribution to the susceptibility should always be positive.

The result (40) can be compared with the analogous one in the spherical model [15]. There the smooth part of the function $\Delta$ at small $\tau$ and $\varepsilon$ is also a constant proportional to $\sqrt{\varepsilon}$. However, there is the family of special cooling trajectories where $\Delta$ have no smooth part and cooling at zero field belongs to this family. It is not clear whether such a family exists in the Ising model.

Consider now anomalous response function at $\tau, \tau' \gg \varepsilon$. The equations for $q$ and $\Delta$ in this region are:

$$\Delta(t,t') \left( \varepsilon \log \frac{q_{t,t'}}{\sqrt{\tau\tau'}} + q_{t,t'}^2 - \tau^2 - \tau'^2 \right) + \int \Delta_{t,1} \Delta_{1,t'} \, d1 = 0 ,$$

$$q_{t,t'} \left( \varepsilon \log \frac{q_{t,t'}}{\sqrt{\tau\tau'}} + \frac{2}{3} q_{t,t'}^2 - \tau^2 - \tau'^2 \right) +$$
$$q_{t,t_c} q_{t',t_c} + \int (q_{t,1} \Delta_{t',1} + q_{t',1} \Delta_{t,1}) \, d1 = 0 ,$$

$$q_{t,t_c} \left( \varepsilon \log \frac{q_{t,t_c}}{\sqrt{\tau\varepsilon}} \right) + \int \Delta_{t,1} q_{1,t_c} \, d1 = 0 . \qquad (41)$$

At $\tau, \tau' \gg \sqrt{\varepsilon}$ the terms containing $\varepsilon$ in (41) are small and solution is closed to one for the Sherrington-Kirkpatrick model:

$$\Delta(t,t') = 2\tau' \frac{d\tau'}{dt'} \theta(t-t'), \ q(t,t') = 2\tau . \qquad (42)$$

In the inverse limit $\tau, \tau' \ll \sqrt{\varepsilon}$ we can introduce the notations $\Delta(t,t') = \sqrt{\frac{\tau\varepsilon}{\tau'}} \Theta(t,t')$ and $q(t,t') = \sqrt{\frac{\tau\tau'}{\varepsilon}} w(t,t')$ and write these equations as

$$\Theta(t,t') w(t,t') + \int \Theta_{t,1} \Theta_{1,t'} \, d1 = 0 ,$$

$$\frac{1}{2} w^2(t,t') + \int (w_{t,1} \Theta_{t',1} + w_{t',1} \Theta_{t,1}) \, d1 + \frac{2}{3} \tau\tau' - \tau^2 - \tau'^2 = 0 . \qquad (43)$$

We have not found analytical solutions of the equations (43), but it seems evident that these solutions (which may be found numerically) interpolate smoothly between the results (40) and (42) corresponding to the regions $(\tau, \tau' \ll \varepsilon)$ and $(\tau, \tau' \gg \sqrt{\varepsilon})$ respectively. In another terms, there is no qualitative difference between the solutions in these two regions. This result looks quite surprising: it was found within static replica approach by Gardner [10] that the second phase transition should exist in this model (at $\tau \sim \sqrt{\varepsilon}$), which is characterized by full replica symmetry breaking. In the dynamic approach we have not found room for such a transition.

## 5.3 Susceptibility and heat capacity

Now it is easy to find observable values such as field cooled and zero field cooled susceptibilities and heat capacity. Let's consider the region $\tau \ll \varepsilon$. The value of $\chi_{FC}$ is

$$\chi_{FC} = \int_{-\infty}^{t} G(t, t') \beta(t') dt' = \frac{1}{T_c}$$

with the accuracy of $\tau \sqrt{\varepsilon}$ and hence have a jump in the derivative with respect to temperature. Zero field cooled susceptibility $\chi_{ZFC}$ is determined by the integral of "fast" part of response function:

$$\chi_{ZFC} = \int_{-\infty}^{t} \left( G(t, t') - \Delta(t, t') \right) \beta(t') dt' = \frac{1 - \varepsilon - \tau}{T_c}$$

and has a jump at the transition point.

Heat capacity $C = dE/dT$ also has a jump, in order to derive it the internal energy should be written in terms of functions $C(t, t')$ and $G(t, t')$ [15]:

$$\beta(t) E(t) = -\int_{-\infty}^{t} C^{p-1}(t, t') G(t, t') dt'$$

Derivative of this equation with respect to temperature yields a downward jump in the heat capacity

$$\Delta C = C_{T_c - 0} - C_{T_c + 0} = -\varepsilon \ .$$

(note that in the standard Landau theory of second-order phase transtions heat capacity shows an upward jump on temperature decrease). Static theory [11] also predicts a jump in heat capacity, but at lower (static transition) temperature.

# 6 Conclusions

We have studied dynamics of the $p = 2 + \varepsilon$-spin interaction Ising spin glass above and below the dynamics transition temperature $T_D$ (implicitly defined by Eq.(21). The discontinuous transition is found at zero and weak external magnetic fields, $b \leq b_{tr} \sim \varepsilon$, whereas at higher fields $b \geq b_{tr}$ the transition is continuous and resembles (though definitely is not identical to) the SK model transition. Near the "tricritical point" $b = b_{cr}$ the dynamic exponent determining the rate of long-time relaxation right at the transition is approaching zero: $\nu_1 \propto \sqrt{b - b_{tr}}$.

In the glassy phase the history dependence is described quantitatively by the anomalous response and correlation functions $\Delta(t, t')$ and $q(t, t')$. We have derived equations for these functions and (in the case of zero external field) solved them in several regions of reduced temperature $\tau \ll 1$. Very close to $T_D$, at $\tau \ll \varepsilon$, the main contribution to the anomalous response function $\Delta(t, t')$ comes from the $\delta$-function term (cf.Eq.(35)), in agreement with replica theory solution [10]. Indeed, the one-step replica symmetry breaking found in [10] is commonly interpreted [8] as a signature of an instantaneous "appearance" of an exponentially large number of valleys right at the transition (i.e. of the extensive configurational entropy $S_{con} \propto N$), those states do not acquire additional "fine structure" (and their number does not grow) as temperature decreases further. Such a structure of equilibrium valleys would precisely agree with the "*delta*-function only" solution for $\Delta(t, t')$: the physical idea behind the *slow cooling* approach is that anomalous response is nonzero when the number of valleys grows (with $T$ decrease usually) making it possible to lower free energy by the proper choice of the valley.

However we found that even in the smallest-$\tau$ range the $\Delta(t, t')$ response contains also the smooth part (cf.Eq.(40)) corresponding, within the same logic, to the continuous splitting of valleys below $T_D$. The same conclusion was reached by a different route in a recent preprint [8]. The above results provide additional (cf. [15]) evidence that the structure of valleys most relevant for slow dynamics is different from those responsible for equilibrium Gibbs partition function (which is reflected already in the fact that the dynamic transition temperature is higher than the static one); for the discussion of the additional aspects of relation between dynamics and static quantities see [22].

Heat capacity and zero field cooled susceptibility have a downward jump at the transition temperature. It should be mentioned that similar jump in heat capacity has been observed experimentally in real liquid-glass transitions [23].

At lower temperatures $\tau \gg \sqrt{\varepsilon}$ the solution of *slow cooling* equations is approaching the one of the SK model [2, 4], whereas in the intermediate region $\varepsilon \ll \tau \ll \sqrt{\varepsilon}$ it interpolates between the above limiting cases. Since qualitative features of the solutions are similar in the regions $\tau \ll \varepsilon$ and $\tau \gg \sqrt{\varepsilon}$, we do not expect any additional phase transitions within *slow cooling* approach. Thus we found that the low-temperature properties of the SK ($p = 2$) and $p = 2 + \varepsilon$ Ising glass models are rather similar, in spite of the drastic difference known to exist between the corresponding phase transitions.

Real experiments on as well as Monte-Carlo simulations of glassy systems are always done within limited time scale, which makes virtually impossible to observe Gibbs equilibrium properties which are the subject of static replica theory; on the other hand, *slow cooling* approach seems to be most suited to describe finite-time experiments. Unfortunately it does not seem possible to compare directly the present analytical results with Monte-Carlo simulation since our calculations are done for small $\varepsilon \ll 1$; however we expect qualitative features of our solution to survive for, e.g. $p = 3$ Ising glass model which could be simulated directly.

Let us note finally that the potentially interested problem which we have not studied here is the anomalous response behavior close to the tricritical point $b = b_{cr}$ (where dynamic exponents $\nu_1$ and $\nu_2$ tend to zero).

# 7 Acknoledgements

We are grateful to P.Chandra and L.Ioffe for many useful comments. This work was supported by the International Science Foundation grant M6M000 and INTAS grant 93-2492 (within research program of International Center for Fundamental Physics in Moscow).

# Appendix 1

Let us derive at first the generating functional $Z[\hat{\lambda}]$ for one Ising spin in the arbitrary field. Distribution function for (5) obeys equation

$$\partial_t P\{\sigma,t\} + \hat{L}P\{\sigma,t\} = 0 , \qquad (44)$$

where

$$\hat{L}P\{\sigma,t\} = -\sum_{\{\sigma'\}} [w\{\sigma \to \sigma'\}P\{\sigma,t\} - w\{\sigma' \to \sigma\}P\{\sigma',t\}]$$

We will be also interesting in the function $P(\sigma,t,\sigma',t')$ — probability to find spin $\sigma$ at the time $t$ provided that at the time $t'$ it's value was $\sigma'$, $P(\sigma,t,\sigma',t') = 0, t < t'$. This probability is a Green's function of the operator $\partial_t + \hat{L}_\sigma$:

$$\left(\partial_t + \hat{L}_\sigma\right) P(\sigma,t,\sigma',t') = \delta(t-t')\delta_{\sigma,\sigma'}$$

Solution of this equation is

$$P(\sigma,t,\sigma',t') = \frac{1}{2}(1 + \sigma m_t(\sigma',t')) \ (t > t') ,$$

where

$$m_t(\sigma',t') = \sigma' exp\,(t'-t) + \int_{t'}^{t} exp\,(\tau - t)\tanh \beta b_\tau \ d\tau.$$

We also can write the solution of (44), i.e distribution function:

$$P(\sigma,t) = \frac{1}{2}(1 + \sigma m_t(m_0,t_0))$$

The process described by the equation (5) is markovian. For this reason correlation function $= \langle \sigma_{t_1} \sigma_{t_2} \cdots \sigma_{t_n} \rangle$ is determined only by distribution and probability functions:

$$C = \sum_{\{\sigma_1 \sigma_2 \cdots \sigma_n\}} \sigma_1 P(\sigma_1,t_1,\sigma_2,t_2)\sigma_2 P(\sigma_2,t_2,\sigma_3,t_3) \times$$
$$\sigma_{n-1}P(\sigma_{n-1},t_{n-1},\sigma_n,t_n)\sigma_n P(\sigma_n,t_n) \qquad (45)$$

for $(t_1 > t_2 > \cdots > t_n > t_0)$.

Later it will be convenient to introduce auxiliary functions

$$A(t) = exp\left(\int_{-\infty}^{t} \hat{\lambda}(\tau)m(\tau)\, d\tau\right)$$

and

$$B(t) = m(t)exp\left(\int_{\infty}^{t} \hat{\lambda}(\tau)m(\tau)\, d\tau\right)$$

where $m(t)$ obeys equation

$$\partial_t m(t) = i\hat{\lambda}(t)(1 - m^2(t)) - (m(t) - \tanh\beta b), \quad m(t_0) = m_0. \qquad (46)$$

This functions are connected by the relation

$$B(t) = \int_{-\infty}^{t} exp\,(\tau - t)(\hat{\lambda}_\tau + \tanh\beta b_\tau)A(\tau)\, d\tau$$

Suppose that

$$\langle \sigma_2 \cdots \sigma_n \rangle = \hat{\delta}_2 \cdots \hat{\delta}_n Z_0 \Big|_{\hat{\lambda}=0} \quad, t_2 > \cdots > t_n\ ,$$

where

$$Z_0[\hat{\lambda}] = \left\langle exp\left(\int_{t_0}^{\infty} \hat{\lambda}(t)\sigma(t)dt\right)\right\rangle = exp\left(\int_{t_0}^{\infty} \hat{\lambda}(t)m(t)dt\right)\ , \qquad (47)$$

and prove the same equation for $n$-point correlator.

On the one hand one can find using (45)

$$\langle \sigma_1 \sigma_2 \cdots \sigma_n \rangle = \langle \sigma_3 \cdots \sigma_n \rangle\, exp(t_2 - t_1)\, +$$
$$+ \langle \sigma_2 \cdots \sigma_n \rangle \int_{t_2}^{t_1} exp\,(\tau - t_1)\tanh\beta b_\tau\, d\tau\ . \qquad (48)$$

On the other hand, for $t > t_1 > \cdots > t_n > t_0$

$$\hat{\delta}_1 \hat{\delta}_2 \cdots \hat{\delta}_n Z_0[\hat{\lambda}]\Big|_{\hat{\lambda}=0} = \hat{\delta}_1 \hat{\delta}_2 \cdots \hat{\delta}_n A(t)\Big|_{\hat{\lambda}=0} =$$

$$\left.\hat{\delta}_2\cdots\hat{\delta}_n B(t_1)\right|_{\hat{\lambda}=0} = \hat{\delta}_3\cdots\hat{\delta}_n\left[exp(t_2-t_1)A(t_2)+\right.$$

$$\left.\left.\int_{t_2}^{t_1} exp(\tau-t_1)(\hat{\lambda}_\tau+\tanh\beta b_\tau)\hat{\delta}_2 A(\tau)\,d\tau\right]\right|_{\hat{\lambda}=0} =$$

$$\left.\hat{\delta}_3\cdots\hat{\delta}_n\left[exp(t_2-t_1)A(t)+\int_{t_2}^{t_1} exp(\tau-t_1)\tanh\beta b_\tau\,\hat{\delta}_2 A(t)\,d\tau\right]\right|_{\hat{\lambda}=0}$$

which is identical to (48).

The procedure of the generalization of the functional $Z_0$ to the case of interacting spins is described in [19, 20]. Let us denote $C = \langle\sigma_1\sigma_2\ldots\sigma_k\rangle$, where indices are the unions of the time and space arguments. Now one can introduce auxiliary fields $h$ and $\hat{h}$ and expand term with interaction over $J$. Correlator $C$ is

$$C = \sum_{n=0}^{\infty}\int\prod_{i=1}^{N}\mathcal{D}h_i\mathcal{D}\hat{h}_i exp\left(\sum_{i=1}^{N}-i\int\hat{h}_i(h_i-\beta b)dt\right)\left\langle\frac{1}{n!}K_n\right\rangle, \qquad (49)$$

where

$$K_n = \sigma_1\sigma_2\ldots\sigma_k\left(\sum_{i_1<i_2<\cdots<i_p}\int dt J_{i_1\ldots i_p} i\hat{h}_{i_1}\sigma_{i_2}\cdots\sigma_{i_p}\right)^n,$$

and $\langle\ldots\rangle$ means average over dynamics of the noninteracting spins in the field $h_i$. $\langle K_n\rangle$ can be written as a variation of $Z_0$:

$$\sum_{n=0}^{\infty}\left\langle\frac{1}{n!}K_n\right\rangle = \frac{1}{n!}\sum_{n=0}^{\infty}\hat{\delta}_1\hat{\delta}_2\ldots\hat{\delta}_k\left(\sum_{i_1<i_2<\cdots<i_p}\int dt J_{i_1\ldots i_p} i\hat{h}_{i_1}\hat{\delta}_{i_2}\cdots\hat{\delta}_{i_p}\right)^n \left.Z_0\right|_{\hat{\lambda}=0} =$$

$$\left.\hat{\delta}_1\hat{\delta}_2\ldots\hat{\delta}_k\,exp\left(\sum_{i_1<i_2<\cdots<i_p}\int dt J_{i_1\ldots i_p} i\hat{h}_{i_1}\hat{\delta}_{i_2}\cdots\hat{\delta}_{i_p}\right)Z_0\right|_{\hat{\lambda}=0}$$

where $\hat{\delta}_t = \delta/\delta\hat{\lambda}(t)$. In the expression (49) the integration over $h$ and $\hat{h}$ can be performed, which replaces $h$ by $\beta b$ and $i\hat{h}$ by $\delta/\delta\beta b$. Thus the generating functional $Z$ is

$$Z = \left.\hat{J}Z_0\right|_{h=\beta b}$$

where $\hat{J}$ is
$$\hat{J} = exp\left(\sum_{i_1<i_2<\cdots<i_p}\int dt J_{i_1\ldots i_p}\hat{\delta}_{i_1}\hat{\delta}_{i_2}\cdots\hat{\delta}_{i_p}\right)$$

## Appendix 2

Let's show how one can derive adiabatic generating functional (26) starting from Langevin dynamics of soft spins:

$$\Gamma_0^{-1}\partial_t\sigma_i(t) = -\frac{\partial\beta H}{\partial\sigma_i(t)} + r_0\sigma_i - u\sigma_i^3 + \xi_i(t) \ . \qquad (50)$$

Here $H$ is the Hamiltonian (3), $\xi_i(t)$ is a white noise with zero mean and variance
$$<\xi_i(t)\xi_j(t')> = 2\Gamma_0^{-1}\delta_{ij}\delta(t-t') \ .$$
The constants $r_0$ and $u$ should tend to infinity in the Ising limit with $r_0/u = 1$.

As was shown in [11, 15], average this Langevin equation over disorder yields

$$\partial_t\sigma(t) = r_0\sigma(t) - u\sigma^3(t) + \int_{-\infty}^{t} dt' \ \hat{G}(t,t')\sigma(t') + h(t) + \xi(t) \qquad (51)$$

with nonlocal noise
$$<\xi(t)\xi(t')> = 2\Gamma_0^{-1}\delta(t-t') + \hat{C}(t,t') \ . \qquad (52)$$

The next step is to divide the functions $C$ and $G$ and noise $\xi$ into slow and fast parts [18, 2]:

$$C(t,t') = \tilde{C}(t-t') + q(t,t'), \ G(t,t') = \tilde{G}(t-t') + \Delta(t,t').$$
$$\xi(t) = \tilde{\xi}(t) + z(t),$$
$$<z(t)z(t')> = \sqrt{\mu(t)\mu(t')}q^{p-1}(t,t') \qquad (53)$$

Integration over fast noise leads to the equation for the slow magnetization [2] which replaces Langevin equation in the adiabatic limit:

$$\langle\sigma\rangle_{\tilde{\xi}} = m(t) = \tanh(H_{eff}(t)) \qquad (54)$$

where in the case of Ising spins

$$H_{eff}(t) = z(t) + \beta b(t) + \int_{-\infty}^{t} dt' \, \hat{\Delta}(t,t') m(t') \, . \tag{55}$$

Now we can easily construct adiabatic generating functional $Z[\hat{\lambda}]$. Note that correlation function can be written as:

$$C(t_1, \ldots, t_n) = \langle m(t_1) \ldots m(t_n) \rangle_z \, .$$

Thus,
$$Z[\hat{\lambda}] = \left\langle \int \mathcal{D}h \mathcal{D}\hat{h} \, J \, exp(S) exp(\int \hat{\lambda} \tanh \beta b) \right\rangle_z \tag{56}$$

where
$$S = -i \int \hat{h}_t (h_t - z_t - \beta b_t - \int \hat{\Delta}_{t,t'} \tanh h_{t'} dt') dt \, ,$$

and Jacobian
$$J = \frac{\partial(h_t - z_t - \beta b_t - \int \hat{\Delta}_{t,t'} \tanh h_{t'} dt')}{\partial h_{t'}} =$$
$$exp\left(-\int \hat{\Delta}_{t,t}(1 - \tanh^2 h_t) dt\right) = 1$$

since we should put $\Delta(t,t) = 0$ in the adiabatic limit, as was mentioned in the previous section. Average over $z$ yields

$$S = -i \int \hat{h}(h - \beta b) - \frac{1}{2} \int \hat{h}_1 \hat{h}_2 \sqrt{\mu_1 \mu_2} q_{1,2}^{p-1} + i \int \hat{h}_1 \tanh h_2 \Delta_{1,2} \, .$$

Then we should expand $Z[\hat{\lambda}]$ over $q$ and $\Delta$ and rewrite each term of the expansion as a variation of $Z_0 = exp(\int \hat{\lambda} \tanh h)$. After some algebra one can derive formula (26).